\documentstyle[prc,aps,preprint,psfig]{revtex}
\draft
\begin{document}
\title{Dilepton production and $m_T$-scaling at BEVALAC/SIS energies
\thanks{Work supported by BMBF and GSI Darmstadt.}}
\author{E. L. Bratkovskaya$^1$, W. Cassing$^1$,
R. Rapp$^2$ and J. Wambach$^3$\\[5mm]
$^1$Institut f\"ur Theoretische Physik,
Universit\"at Giessen\\D-35392 Giessen, Germany \\
$^2$ Department of Physics and Astronomy, State University of New York, \\
Stony Brook, NY 11794-3800, U.S.A. \\
$^3$ Insitut f\"ur Kernphysik, TU Darmstadt, Schlo\ss gartenstr. 9,\\
 D-64289 Darmstadt, Germany}
\maketitle

\begin{abstract}
We present a dynamical study of $e^+e^-$ production in C + C and Ca +
Ca collisions at BEVALAC/SIS energies on the basis of the covariant
transport approach HSD employing momentum-dependent $\rho$-meson
spectral functions that include the pion modifications in the nuclear
medium as well as the polarization of the $\rho$-meson due to resonant
$\rho$$-N$ scattering.  We find that the experimental data from the DLS
collaboration cannot be described  within the $\rho$-meson spectral
function approach.  A dropping $\eta$-mass scenario leads to a good
reproduction of the DLS dilepton data, however, violates the
$m_T$-scaling of $\pi^0$ and $\eta$ spectra as observed by the TAPS
collaboration as well as $\eta$ photoproduction on nuclei.
\end{abstract}

\noindent
PACS: 25.75+r \ 14.60.-z \  14.60.Cd

\noindent
Keywords: relativistic heavy-ion collisions, leptons

\narrowtext
\section{Introduction}

The mechanism of chiral symmetry restoration at high baryon density is
of fundamental importance for the understanding of
QCD~\cite{brownrho,chiral}, but a clear experimental evidence has not
been achieved, so far.  In this respect, dileptons are particularly
well suited for an investigation of the early phases of high-energy
heavy-ion collisions because they can leave the reaction volume
essentially undistorted by final-state interactions.  Indeed,
dileptons from heavy-ion collisions have been measured by the DLS
collaboration at the BEVALAC~\cite{ro88,na89,ro89,DLSnew} and by the
CERES~\cite{CERES,Ullrich}, HELIOS~\cite{HELIOS,HELI2},
NA38~\cite{NA38} and NA50~\cite{NA50} collaborations at SPS energies.

Various theoretical calculations
\cite{Li,Cass95C,Koch96,Srivas,Frankfurt} fail in describing the
spectral shape of the dilepton yield for nucleus-nucleus collisions at
SPS energies when employing a 'free' $\rho$-meson form factor but are
successful when using a dropping $\rho$-mass in the medium.  However, a
more conventional approach including the change of the $\rho$-meson
spectral function in the medium due to the coupling of $\rho, \pi,
\Delta$ and nucleon dynamics along the lines of
Refs.~\cite{Herrmann,asakawa,Chanfray,Rapp} was found to be (roughly)
compatible with the CERES data~\cite{Cass95C,Rapp} as well, whereas
the dimuon data of the HELIOS-3 collaboration~\cite{HELIOS} could only
be described satisfactorily when including 'dropping' meson
masses~\cite{Li96,Cass96H} according to Hatsuda and Lee~\cite{hatsuda}.
Meanwhile, our knowledge on the $\rho$ spectral function has improved
since -- as first pointed out by Friman and Pirner~\cite{Friman} --
resonant $\rho$-$N$ interactions in p-wave scattering significantly
enhance the strength in the vector-isovector channel at low invariant
mass. In fact, the CERES data for S~+~Au at 200~A~GeV and Pb~+~Au at
160~A~GeV, using an expanding fireball model, were found to be
compatible within such a hadronic scenario~\cite{RappNPA}.
This general shape of the $\rho$ spectral function
has also been confirmed recently in Ref. \cite{Peters} where
the authors find a significant smearing of the $\rho$ strength at low
$\rho$ momenta due to a selfconsistent evaluation of the resonance
widths. Furthermore,
as shown in Refs.~\cite{Weise97,Leupold}, both scenarios (dropping mass and
$\rho$-meson broadening) are compatible with QCD sum rules.

The thermodynamical result of Ref. \cite{RappNPA} was supported by
HSD transport calculations \cite{CBRW97} where an improved $\rho$
spectral function ({\it i.e.}~by constraints from $\gamma p$ and
$\gamma A$ data) was involved. The important conclusion in
Ref.~\cite{CBRW97} was  that both the 'dropping mass'
scenario~\cite{Cass95C,Li96,Cass96H,Brat97} as well as the hadronic
spectral function approach~\cite{RappNPA} lead to dilepton spectra
which are in good agreement with the experimental data for all systems
at SPS energies.

In this respect, it is a natural step to investigate within the same
framework dilepton production at BEVALAC energies where quite a
different temperature and density regime is probed. From a recent
publication of the DLS collaboration~\cite{DLSnew}, it turns out that
the new measurements of dilepton yields in A~+~A collisions have
increased by about a factor of 6-7 in comparison to the previous data
set~\cite{ro88,na89,ro89} due to an improvement of the DLS detector and
the data analysis correcting for dead time losses. However, the first
generation of DLS data were reasonably described by several
theoretical models~\cite{Xiong90,Wolf90,Gudima,BCMas96} without
incorporating any medium effects.  These various models are consistent
in the general tendency that the dilepton pairs of invariant mass below
0.4 GeV are primarily produced from hadronic sources, such as the
Dalitz decays of $\pi^0, \Delta$ and $\eta$ and $pn$ bremsstrahlung,
whereas for $M \ge 0.4$~GeV the dominant contributions stem from the
$\pi^+\pi^-$ annihilation channel and direct decays of vector mesons
($\rho,\omega$).  Since this has lead to a fair agreement with the old
DLS data the aforementioned theoretical models are not able to explain
the new DLS data set~\cite{DLSnew} without incorporating  new
ingredients such as vector meson selfenergies.

The DLS collaboration pointed out that the inconsistency between the
data and theoretical results -- incorporating no medium effects --
might be due to a strong underestimation
of the $\eta$ contribution in the model calculations.  As found in
Ref.~\cite{DLSnew} the shape of the dilepton yield for $M\le 0.4$~GeV
is quite similar to that of the $\eta$ component in the BUU
model~\cite{BCMas96}. The main knowledge about $\eta$ production in
heavy-ion collisions at BEVALAC/SIS energies is provided by the TAPS
collaboration~\cite{TAPSold,TAPS-CC,TAPS-CaNi}. However, TAPS measures
$\pi^0$ and $\eta$ mesons in a small rapidity interval near $y_{cm}$
whereas the DLS spectrometer covers the  forward rapidity region.
Within a thermal model calculation -- using the TAPS mid-rapidity data
and assuming an isotropic $\eta$ angular distribution -- the DLS collaboration
found that the $\eta$ yield in their rapidity
window then is in good agreement with the BUU model
predictions~\cite{BCMas96,Teiseta}. The dilepton enhancement observed by
the DLS Collaboration thus might be due to a strong anisotropy of the
$\eta$ angular distribution in the center-of-mass system.

In this article we address the DLS 'puzzle' by employing in our
transport calculation  novel theoretical developments, i.e. in-medium
$\rho$ spectral functions as well as hypothetical
in-medium modifications of the $\eta$
mass.  To arrive at more stringent conclusions concerning the $\eta$
contribution we simultaneously analyze the DLS dilepton
data~\cite{DLSnew} together with the new TAPS
data~\cite{TAPS-CC,TAPS-CaNi} with respect to $m_T$ spectra of
$\pi^0$'s and $\eta$'s as well as $\eta$ photoproduction on nuclei.

Our paper is organized as follows:  In Section~2 we briefly describe
the transport approach employed as well as the elementary production
channels.  Section~3 contains a comparison of our calculations for
$\pi^0$ and $\eta$ $m_T$-spectra with the TAPS data. In Section~4 we
discuss the $\rho$ spectral functions which will be applied for
calculating the dilepton spectra.  Section~5 contains a detailed
comparison of the calculated dilepton spectra at BEVALAC energies with
the DLS data~\cite{DLSnew}. In Section~6 we investigate the
consequences of a dropping $\eta$-mass scenario for $m_T$-spectra and
dilepton production at BEVALAC/SIS energies as well as $\eta$
photoproduction on nuclei. The $\eta$
angular anisotropy is investigated in Section~7 while we close
with a summary and discussion of open problems in Section~8.

\section{Transport Model ingredients}

In this work we perform the theoretical analysis along the line of the
HSD approach \cite{Ehehalt} which is based on a coupled set of
covariant transport equations for the phase-space distributions $f_{h}
(x,p)$ of hadron $h$ \cite{Ehehalt,Weber1}, i.e.
\begin{eqnarray}  \label{g24}
\lefteqn{\left\{ \left( \Pi_{\mu}-\Pi_{\nu}\partial_{\mu}^p U_{h}^{\nu}
-M_{h}^*\partial^p_{\mu} U_{h}^{S} \right)\partial_x^{\mu}
+ \left( \Pi_{\nu} \partial^x_{\mu} U^{\nu}_{h}+
M^*_{h} \partial^x_{\mu}U^{S}_{h}\right) \partial^{\mu}_p
\right\} f_{h}(x,p) } \nonumber \\
&& = \sum_{h_2 h_3 h_4\ldots} \int d2 d3 d4 \ldots
 [G^{\dagger}G]_{12\to 34\ldots}
\delta^4_\Gamma(\Pi +\Pi_2-\Pi_3-\Pi_4 \ldots )  \nonumber\\
&& \times \left\{ f_{h_3}(x,p_3)f_{h_4}(x,p_4)\bar{f}_{h}(x,p)
\bar{f}_{h_2}(x,p_2)\right.  \nonumber\\
&& -\left. f_{h}(x,p)f_{h_2}(x,p_2)\bar{f}_{h_3}(x,p_3)
\bar{f}_{h_4}(x,p_4) \right\} \ldots\ \ .
\end{eqnarray}
In Eq.~(\ref{g24}) $U_{h}^{S}(x,p)$ and $U_{h}^{\mu}(x,p)$ denote the
real part of the scalar and vector hadron selfenergies, respectively, while
$[G^+G]_{12\to 34\ldots} \delta^4_\Gamma (\Pi+\Pi_2-\Pi_3-\Pi_4\ldots )$
is the `transition rate' for the process $1+2\to 3+4+\ldots$.
In quantum many-body systems this transition rate is partly off-shell --
as indicated by the index $\Gamma$ of the $\delta$-function --
with width $\Gamma \approx \tau_{coll}^{-1}$ where $\tau_{coll}$ is the
average collision rate for the colliding hadrons.
In our semi-classical transport approach the on-shell limit $\Gamma \to
0$ is adopted which is found to describe reasonably well hadronic
spectra in a wide dynamical regime. Note that the on-shell limit has
been proven explicitly to be a valid approximation for intermediate
energy heavy-ion collisions in Ref.~\cite{Cassing87}. The hadron
quasi-particle properties in (\ref{g24}) are defined via the mass-shell
constraint~\cite{Weber1},
$\delta (\Pi_{\mu}\Pi^{\mu}-M_{h}^{*2})$, with effective masses
and momenta given by
\begin{eqnarray} \label{g26}
M_{h}^* (x,p)&=&M_h + U_h^{{S}}(x,p) \nonumber \\
\Pi^{\mu} (x,p)&=&p^{\mu}-U^{\mu}_h (x,p)\ \ ,
\end{eqnarray}
while the phase-space factors
\begin{equation}
\bar{f}_{h} (x,p)=1 \pm f_{{h}} (x,p)
\end{equation}
account for fermion Pauli-blocking or Bose enhancement,
depending on the type of hadron in the final and initial
channel. The dots in Eq.~(\ref{g24}) indicate further contributions
to the collision term with more than two hadrons in the final/initial
channels. The transport approach (\ref{g24}) is fully specified by
$U_{h}^{S}(x,p)$ and $U_{h}^{\mu}(x,p)$ $(\mu =0,1,2,3)$, which
determine the mean-field propagation of the hadrons, and by the
transition rates $G^\dagger G\,\delta^4_\Gamma (\ldots )$ in the collision
term that describe the scattering and hadron production/absorption
rates. For more details we refer the reader to Ref.~\cite{Ehehalt}.

In addition to the direct production of all mesons in
Ref.~\cite{Ehehalt} we treat the  production of vector mesons
($\rho,\omega,\phi$) and $\eta$'s at low energies also perturbatively.
The vector mesons as well as $\eta$ mesons are produced in
baryon-baryon and meson-baryon collisions according to the elementary
production cross sections evaluated in Refs.~\cite{Sib97,Vetter} on
the basis of boson-exchange diagrams. For an explicit description of the
parametrizations employed we refer the reader to Eqs.~(14) and (18) and
Tables 1-3 in Ref.~\cite{Sib97} for $\rho, \omega$ and $\phi$ production
in pion-baryon and baryon-baryon channels.
The $\eta$ cross section from $pp$ collisions is taken in line with
the boson exchange calculation in Ref. \cite{Vetter}, which can be fitted as
\begin{equation}
\label{vetter}
\sigma_{pp \to pp \eta}(\sqrt{s}) = a \ {\sqrt{s} - \sqrt{s_0}\over
b + (\sqrt{s}-\sqrt{s_0})^2}
\end{equation}
with $a = 0.4$ mb  and $b$= 0.552 GeV$^2$ while $\sqrt{s_0}$ is the
$\eta$ production threshold. This parametrization also reproduces the
$\eta$ production data from the PINOT Collaboration \cite{Pinot} rather
well.  The
$pn \to pn\eta$ cross section is taken to be about 6 times  the $pp \to
pp\eta$ cross section close to threshold in line with the new data from
the WASA collaboration~\cite{WASA}.  The production of $\eta$ mesons in
pion-nucleon collisions as well as $\eta$ scattering and reabsorption
on nucleons is taken from Ref. \cite{Cassing91} employing the
experimental data from \cite{Landolt} and using detailed balance as
well as the known branching ratios of the $N(1535)$ to pions and
$\eta$-mesons. The respective $\eta$ absorption cross section was found
in Ref. \cite{Roebig} (Fig. 2) to be well in line with the experimental
$\eta$ photoproduction data on nuclei (see below).

Apart from the production cross sections of the vector mesons $\rho$ and
$\omega$ their final state interactions with nucleons have to be considered.
For the $\rho$ we adopt the resonance coupling model specified in Section 2 of
Ref. \cite{SibCas} whereas for the $\omega$ we use the total and inelastic
cross sections (Eqs. (34), (35)) of Ref. \cite{Golubeva} which are less
constraint by experimental data. The latter uncertainty, however, plays no
essential role in the present analysis because the $\omega$-mesons dominantly
decay in the vacuum and thus do not contribute to the dilepton invariant
mass spectrum below 0.6 GeV.

As noted above, the transport approach (HSD) was found to describe
reasonably well hadronic data at SIS energies as well as hadronic and
dilepton data at SPS energies~\cite{Cass95C,Cass96H,Brat97,Ehehalt}.

\section{$m_T$ scaling}

In relativistic heavy-ion collisions at BEVALAC/SIS energies the nuclei
can be compressed up to about 3 times normal nuclear matter density
$\rho_0$.  In the hot compression zone the nucleons are exited to
baryonic resonances which decay by emitting mesons and the produced
mesons then can be absorbed, re-emitted and re-scattered.

Information about the degree of equilibration can be obtained from
$m_T$ spectra, where $m_T=(p_T^2+m^2)^{1/2}$ and $p_T$ is the
transverse momentum of a particle with mass $m$. In this case the
spectrum $1/m_T^2 d\sigma/d m_T$ is of Boltzmann type $\sim \exp(-\beta
m_T)$ with a slope parameter $\beta$ being related to the global
(inverse) temperature at freeze-out in the absence of flow.

The transverse-mass spectra of $\pi^0$ and $\eta$ mesons in heavy-ion
collisions at SIS energies were recently measured by the TAPS
collaboration~\cite{TAPSold,TAPS-CC,TAPS-CaNi} where a $m_T$ scaling
has been observed for both mesons.  Such a universal property of the
meson spectra at SIS energies has been predicted  by the Quark Gluon
String Model calculations in Ref.~\cite{Toneev} for the Ar~+~Ca system.
In the remainder of this section we will thus systematically investigate $m_T$
spectra for nucleus-nucleus collisions at SIS energies within the
transport approach specified above.

In Fig.~\ref{Figmt} we compare the results of our calculation for the
transverse-mass spectra of $\pi^0$ and $\eta$ mesons  with the TAPS
data. The upper part shows the $m_T$ spectra for $\pi^0$'s (dashed
histogram) and $\eta$'s (solid histogram) for C~+~C at 1.0~A$\cdot$GeV
in the rapidity interval $0.42 \le y \le 0.74$ and at 2.0~A$\cdot$GeV
for $0.8 \le y \le 1.08$.  The experimental data -- the open circles
and solid squares correspond to $\pi^0$ and $\eta$ mesons,
respectively -- are taken from Ref.~\cite{TAPS-CC}. The theoretical
results as well as the experimental data at 2.0 A$\cdot$GeV  here are
multiplied by a factor of $10^2$.  The middle part corresponds to
Ca~+~Ca at 1.0 A$\cdot$GeV  for $0.48 \le y \le 0.88$ (multiplied by
$10^{-1}$) and at 2.0 A$\cdot$GeV  for $0.8 \le y \le 1.1$ in
comparison with the data from Ref.~\cite{TAPS-CaNi}.  The lower part
shows the calculated $m_T$ spectra for Ni~+~Ni at 1.93~A$\cdot$GeV for
$0.8 \le y \le 1.1$ in comparison with the data from
Ref.~\cite{TAPS-CaNi}.

As seen from Fig.~\ref{Figmt}, the HSD transport model gives a
reasonable description of the $m_T$ spectra of pions and $\eta$'s as
measured by the TAPS collaboration without incorporating any medium
modifications for both mesons. It is important to point out that these
calculations are parameter-free in the sense that all production cross
sections for $\eta$ mesons are extracted from
experimental data in the vacuum and the
$\eta$-nucleon elastic and inelastic cross sections are obtained by using
detailed balance on the basis of an intermediate N(1535) resonance.

\section{In-Medium $\rho$ Propagator}

The dilepton production is calculated perturbatively by including the
contributions from the Dalitz-decays $\Delta\to N l^+l^-$, $N^*\to N
l^+l^-$, $\pi^0 \to \gamma l^+l^-$, $\eta \to \gamma l^+l^-$, $\omega
\to \pi^0 l^+l^-$ and the direct dilepton decays of the vector mesons
$\rho, \omega$ and $\phi$, where the $\rho$ and $\phi$ meson may
be produced in $\pi^+\pi^-$ and $K \bar{K}$ or $\pi \rho$
collisions, respectively.  For a detailed description of $\Delta, N^*$
Dalitz decays we refer the reader to  Ref.~\cite{Wolf90}; the dilepton
decays of the $\eta$ and vector mesons ($\omega, \phi$)
are described in Ref.~\cite{Brat97}.

The dilepton radiation from $\rho$ mesons is calculated as
\begin{equation}
\label{rho}
{dN_{l^+l^-}\over dM} = - Br \ {2M\over \pi} \
{\rm Im}D_\rho (q_0, q; \rho_B, T),
\end{equation}
where $D_\rho$ is the $\rho$-meson propagator in the hadronic medium
depending on the baryon density $\rho_B$ and temperature $T$ as well as
on energy $q_0$ and 3-momentum $q\equiv |\vec q|$ in the local rest
frame of the baryon current ('comoving' frame) while $Br\simeq
4.5\times 10^{-5}$ is the experimental branching ratio to dileptons.
The invariant mass $M$ is related to the $\rho$-meson 4-momentum in the
nuclear medium as
\begin{equation}
\label{m}
M^2 = q_0^2 - q^2 \ ,
\end{equation}
while
\begin{equation}
\label{drho}
{\rm Im} D_\rho = \frac{1}{3} ( {\rm Im} D^L_\rho + 2 {\rm Im} D^T_\rho)
\end{equation}
is spin averaged over the longitudinal and transverse part of the
$\rho$-propagator,
\begin{equation}
\label{Imd}
{\rm Im} D^{L,T}_\rho (q_0, q; \rho_B, T) = {{\rm Im}
\Sigma^{L,T}_\rho (q_0, q; \rho_B, T)\over |M^2 -
\left(m^{(0)}_\rho\right)^2 -
\Sigma^{L,T}_\rho (q_0, q; \rho_B, T)|^2} .
\end{equation}
Following Refs.~\cite{RappNPA,RUBW} the $\rho$-meson selfenergy
$\Sigma^{L,T}_\rho (q_0, q, \rho_B, T)$ is
obtained by combining the effects of the different hadronic interactions:
\begin{equation}
\label{sigrho}
\Sigma^{L,T}_\rho = \Sigma^{L,T}_{\rho\pi\pi}
+ \Sigma^{L,T}_{\rho B B^{-1}} + \Sigma^{L,T}_{\rho\pi a_1} +
\Sigma^{L,T}_{\rho K K_1} .
\end{equation}
The explicit evaluation of the various selfenergy components is
discussed in detail in Ref.~\cite{RappNPA}. We here employ an improved
version in the following respects:
\begin{itemize}
\item[(a)] in addition to p-wave $\rho N \to B$ interactions ($B$=$N$,
$\Delta$, $N(1720)$ and $\Delta(1905)$) we also account for s-wave
excitations into $N(1520)$, $\Delta(1620)$ and $\Delta(1700)$
resonances (the corresponding coupling constants are, as usual,
estimated from the $\rho$$N$ partial decay width);
\item[(b)] the coupling of the resonances to (virtual) photons is
calculated within the improved vector dominance model of Kroll, Lee and
Zumino~\cite{KLZ} to avoid overestimates of the $B\to N\gamma$
branching ratios. This leads to a modification in the coupling of  the
transverse part of the $\rho$ propagator entering
Eqs.~(\ref{rho}),~(\ref{pipian}) such that the combination
$(m_\rho^{(0)})^4\;{\rm Im}D_\rho^T (q_0, q; \rho_B, T)$ is replaced by
the 'transition form factor'
\begin{eqnarray}
\bar{F}^T (q_0, q; \rho_B) & = & -{\rm Im}[\Sigma^T_{\rho\pi\pi}
+\Sigma^{T}_{\rho\pi a_1}+\Sigma^{T}_{\rho K K_1}] \ |d_\rho-1|^2
-{\rm Im}\Sigma^T_{\rho BB^{-1}} \ |d_\rho-r_B|^2
\nonumber\\
d_\rho (q_0, q; \rho_B) & = &
\frac{ M^2-\Sigma^T_{\rho\pi\pi}-\Sigma^{T}_{\rho\pi a_1}
-\Sigma^{T}_{\rho K K_1}- r_B \Sigma^T_{\rho BB^{-1}} }
{ M^2-(m_\rho^{(0)})^2-\Sigma^T_{\rho} } \ ,
\label{KLZ}
\end{eqnarray}
where
\begin{eqnarray}
r_B={ \mu_B \over   {f_{\rho BN}\over m_\rho} \
{(m_\rho^{(0)})^2\over g}}
\nonumber
\end{eqnarray}
denotes the ratio of the photon coupling to its value in the naive VDM.
\item[(c)] the medium modifications of the two-pion selfenergy
$\Sigma_{\rho\pi\pi}$ are extended to arbitrary 3-momentum within the
recently developed model of Urban {\it et al.}~\cite{Urban}.
\end{itemize}
The combined $\rho$-meson selfenergy is then further constrained by
experimental information on $\gamma p$ and $\gamma A$ absorption cross
sections as described in Ref.~\cite{RUBW}. With the aforementioned
improvements (a) to (c) a satisfactory description of the
photoabsorption data, which represent the $M^2 \to 0$ limit of the dilepton
regime, can be achieved on protons as well as on nuclei. In fact, the
melting of the resonance structures (above the $\Delta$ mass), as seen
in the photoabsorption data on nuclei \cite{gammaA}, can be
explained by the broadening of the higher resonances due to a strong
coupling to the short-lived $\rho$-meson in the medium \cite{RUBW}.
This provides a proper basis to assess the
dilepton radiation originating from in-medium $\pi\pi$ annihilation and
$\rho$ decays in A~+~A collisions at BEVALAC/SIS energies.

As an example -- relevant for the heavy-ion reactions at BEVALAC/SIS
energies -- we show in Fig.~\ref{FigRapp} the spin averaged ${\rm - Im}
D_\rho (q_0, q; \rho_B, T)$ in the approach of Ref.~\cite{RappNPA}
(including the modifications (a) to (c)) as a function of the invariant
mass $M$ and the 3-momentum $q$ for a temperature of 70 MeV at $\rho_B
= 0, \rho_0, 2 \rho_0$ and $3 \rho_0$, respectively (note that at this
temperature about 10\% of the nucleons are excited into $\Delta$'s such
that $\rho_N\simeq 0.9\rho_B$). With increasing baryon density we find
the $\rho$ spectral function to increase substantially in width showing
only minor structures at high density. Thus the lifetime of the
$\rho$-meson in the nuclear medium becomes very short due to well
established hadronic interactions.

Furthermore, we will also employ the $\rho$ spectral function from
Ref.~\cite{Peters} which is calculated by taking only the
resonance-holes states (without the correction Eq.~(\ref{KLZ})) into
account including, however, all nucleon resonances up to a mass of
$\approx$ 1.9 GeV.  In addition, in the latter calculation the feedback
of the in-medium $\rho$ spectral function on the resonance widths is
taken into account selfconsistently, i.e. iterated out at given nuclear
matter density and 3-momentum. In Fig.~\ref{FigPeters} we show for
comparison the spectral function from Ref.~\cite{Peters} for $\rho = 0,
\rho_0, 2 \rho_0$ and $3 \rho_0$ at temperature T = 0. While the
overall structure is similar to that in Fig.~\ref{FigRapp} the spectral
functions in the approach of Ref.~\cite{RappNPA} are smoother at low
invariant mass and the structures at small momenta $q$, which result from
a selfconsistent treatment of the resonance widths in
Ref.~\cite{Peters}, are washed out.

The cross section for the pion annihilation channel
$\pi^+ \pi^- \rightarrow \rho^0 \rightarrow e^+ e^-$ is
taken in line with Ref.~\cite{PKoch93} as
\begin{eqnarray}
\sigma_{\pi^+\pi^-\to e^+e^-}(M) = - {16\pi^2\alpha^2\over g_{\rho\pi\pi}^2} \
 {1\over k^2 M^2} \ (m_\rho^{(0)})^4 \ {\rm Im} D_\rho (q_0, q; \rho_B, T)
\label{pipian}
\end{eqnarray}
where $k=(M^2-4m_\pi^2)^{1/2}/2$ is the pion 3-momentum in the
center-of-mass frame and $\alpha$ is the fine structure constant. The
$\rho\pi\pi$ coupling constant and bare $\rho$ mass $m_\rho^{(0)}$ are
fixed to reproduce p-wave $\pi\pi$ scattering and the pion
electromagnetic form factor in free space~\cite{RappNPA,Peters}.

Our dynamical calculations are carried out as follows: the
time-evolution of the proton-nucleus or nucleus-nucleus collision is
described within the covariant transport approach
HSD~\cite{Ehehalt,Brat97} without any dropping vector meson masses.
Whenever a $\rho$-meson is produced in the course of the hadronic
cascade (by baryon-baryon, meson-baryon or pion-pion collisions), its
4-momentum in the local rest frame of the baryon current is recorded
together with the information on the local baryon density, the local
'transverse' temperature  and  its production source. We note that the
definition of a local temperature is model dependent; here we have used
a logarithmic fit to the transverse $p_t$ spectra of mesons at
midrapidity.  Without going into a detailed discussion of this issue we
note that our results for dilepton spectra do not change within the
numerical accuracy when using a constant temperature $T = 70$~MeV at
BEVALAC energies since ${\rm Im} D_\rho$ depends rather weakly on $T$
(at fixed nucleon density).

\section{Dilepton spectra in comparison to BEVALAC data}

In  Fig.~\ref{FigBVfr} we present the calculated dilepton invariant
mass spectra $d\sigma/dM$ for Ca + Ca (upper part) and C + C (lower
part) at the bombarding energy of 1.0 A$\cdot$GeV integrated over
impact parameter and compare them with the new experimental data of the
DLS collaboration~\cite{DLSnew} including the DLS acceptance filter
(version 4.1) as well as a mass resolution $\Delta M/M=10\%$.  The thin
lines indicate the individual contributions from the different
production channels; {\it i.e.}~ starting from low $M$: Dalitz decay
$\pi^0 \to \gamma e^+ e^-$ (dashed line), $\eta \to \gamma e^+ e^-$
(dotted line), $\Delta \to N e^+ e^-$ (dashed line), $\omega \to \pi^0
e^+ e^-$ (dot-dashed line), $N^* \to N e^+ e^-$ (dotted line),
proton-neutron bremsstrahlung (dot-dashed line), $\pi N$ bremsstrahlung
(dot-dot-dashed line); for $M \approx $ 0.8 GeV:  $\omega \to e^+e^-$
(dot-dashed line), $\rho^0 \to e^+e^-$ (dashed line), $\pi^+ \pi^- \to
\rho \to e^+e^-$ (dot-dashed line).  The pion annihilation channel as
well as the direct decay of the vector mesons here have been calculated
with the 'free' $\rho$ spectral function.  The full solid line
represents the sum of all sources.

Our present result is consistent with that from Ref.~\cite{BCMas96} at low
invariant mass $M\le 0.4$~GeV. However, the contributions from direct
decays of $\rho$ and $\omega$ mesons are somewhat larger because novel
elementary production cross sections for pion-baryon collisions from
Ref.~\cite{Sib97} were taken into account as well as $\rho$ and
$\omega$ elastic and inelastic scattering processes with nucleons using
the cross sections from Ref.~\cite{SibCas,Golubeva}.  As seen from
Fig.~\ref{FigBVfr} the new BEVALAC data cannot be properly described in
terms of a 'free' $\rho$ spectral function.  The discrepancy between
the data and the calculations for Ca+Ca as well as for C+C at $0.15\le
M\le 0.4$~GeV is about a factor of 3-5.  At $M\sim m_\rho$ the
calculation is within the error bars except for the last experimental
point.

The description of the DLS data is slightly improved when including the
$\rho$ spectral functions from Ref.~\cite{RappNPA} (cf.
Fig.~\ref{FigRapp}) or Ref.~\cite{Peters} (cf. Fig.~\ref{FigPeters}).
In Fig.~\ref{FigBVsf} we show the dilepton spectra (full solid line)
for Ca~+~Ca (upper part) and C~+~C (lower part) at 1.0~A$\cdot$GeV
calculated with the $\rho$ spectral function from Ref.~\cite{RappNPA}
(Fig.~\ref{FigRapp}) in comparison with the data~\cite{DLSnew}.  The
same comparison is carried out in Fig.~\ref{figp} for the spectral
function (at T=0) from Ref.~\cite{Peters} (Fig.~\ref{FigPeters}).  The
dashed lines correspond to the channel $\rho^0 \to e^+e^-$ while the
dot-dashed lines indicate the contribution from $\pi^+ \pi^- \to \rho
\to e^+e^-$ as described in Section 4. The full $\rho$ spectral
functions lead to a shift of the pion annihilation contribution to the
small invariant mass region as well as to a broadening of the
$\rho$-meson contribution in both approaches. We note that in our
transport approach the pions are on-shell, thus we can not probe the
$\rho$-spectral function in the reactions $\pi^+\pi^- \to \rho^0 \to
e^+e^-$ below the two pion threshold at $M < 2 m_\pi$.  The tails from
the $\pi^+ \pi^-$ annihilation and $\rho$ meson at $M < 2 m_\pi$, which
are seen in Figs.~\ref{FigBVsf} and \ref{figp}, exist only due to the
finite mass resolution of $\Delta M /M \simeq 10\%$.  We note that we
discard meson-baryon ($mB$) and meson-meson ($mm$) bremsstrahlung
channels as well as the Dalitz decays of the baryon resonances
(stemming from secondary pion induced reactions) in order to avoid
double counting when employing the full $\rho$ spectral functions.

Due to the on-shell treatment of pions in the transport approach the
dilepton spectra for $M\simeq 2 m_\pi$ might not be properly described.
In order to investigate the effects from an off-shell propagation of
pions in the channel $\pi^+\pi^- \to \rho^0 \to l^+l^-$, we have
performed additional calculations within the thermodynamical approach
used in Ref.~\cite{RappNPA} at SPS energies.  For BEVALAC energies we
use exactly the same procedure, however, with temperature and density
profiles obtained from the HSD transport model \cite{Ehehalt} for
Ca~+~Ca at 1.0 A$\cdot$GeV.  The solid curves in Fig.~\ref{FigBVtermo}
correspond to the sum of all channels and the $\rho$ decay (from
$\pi^+\pi^-$ annihilation and direct production), respectively, while
the dashed lines are from the transport model (cf. Fig.~\ref{FigBVsf}).
As seen from Fig.~\ref{FigBVtermo} an off-shell pion propagation leads
to a large enhancement in the $\rho$-decay channel especially below the
$2\pi$ threshold. However, considering all channels simultaneously, the
maximum increase -- compared to the transport model -- is only 30\%.
Thus, all independent calculations give practically the same results;
the new DLS data are underestimated at least by a factor of 3 for
$0.15\le M\le 0.4$~GeV even when taking into account the modification
of the $\rho$ meson properties in the nuclear medium.

\section{Dropping $\eta$ mass?}

According to the general scaling idea of Brown and Rho~\cite{BrownRho}
$\eta$ mesons might also change their properties in the medium.  In
this section we examine the possibility of a dropping $\eta$ mass at
BEVALAC/SIS energies as well as for $\eta$ photoproduction on nuclei.
For our analysis we use a linear extrapolation of the $\eta$ mass,
\begin{equation} m^*_\eta = m_\eta^0 \left(1 -
\alpha \frac{\rho_B}{\rho_0}\right),
\label{etamass} \end{equation}
with a 'large' value for $\alpha \approx $ 0.18, which might be
considered as an upper limit.  According to
Eq.~(\ref{etamass}) the $\eta$ mass decreases at normal nuclear matter
density by 18\%. The maximum baryon density reached in central Ca~+~Ca
collisions is about $2.5 \rho_0$, thus the dropping $\eta$ mass leads
to an essential reduction of the $\eta$ production threshold in $BB$
and $mB$ collisions and to a strong enhancement of the $\eta$
population in heavy-ion collisions. Similar, but less pronounced effects
are expected for the $\eta$ photoproduction on nuclei for densities
$\rho_B \leq \rho_0$.

\subsection{Nucleus-nucleus collisions}

In Fig.~\ref{FigBVet} we show the dilepton spectra (full solid line)
for Ca~+~Ca (upper part) and C~+~C (lower part) at 1.0~A$\cdot$GeV
calculated with the full $\rho$ spectral function \cite{RappNPA} and
a dropping $\eta$ mass in comparison with the data from
Ref.~\protect\cite{DLSnew}. As seen from Fig.~\ref{FigBVet}, the
dropping $\eta$ mass scenario together with the full spectral function
approach leads to a good reproduction of the BEVALAC data.

However, the dropping $\eta$ mass scenario does not yield the observed
$m_T$ scaling.  The transverse-mass spectra of $\pi^0$ and $\eta$
mesons for C~+~C in $0.42 \le y \le 0.74$ (lower part) and Ca~+~Ca for
$0.48 \le y \le 0.88$ (upper part) at 1.0~A$\cdot$GeV are shown in
Fig.~\ref{Figmted}. The assignment of lines is the same as in
Fig.~\ref{Figmt}.  The dot-dashed histograms correspond to the $\eta$
$m_T$-spectra obtained within the dropping $\eta$ mass scenario. The
$m_T$ scaling in this case is violated by about a factor of 3
especially at low $m_T \simeq m_\eta$.

Thus we find that the simple $\eta$ dropping mass scheme is not
consistent with the $m_T$ scaling observed by the TAPS collaboration.
A more sophisticated scenario for the $\eta$ self-energy (like momentum
dependent $\eta$ potentials etc. \cite{Oset}) still has to be studied.

\subsection{$\eta$ photoproduction on nuclei}

Since the dynamics of nucleus-nucleus collisions at SIS energies is
rather complex it is useful to have further constraints on the $\eta$
properties in the medium as e.g. tested in $\eta$ photoproduction on
nuclei where the kinematical conditions are much better under control.
The latter problem has been addressed in detail in
Refs.~\cite{Hombach,Effe} on the basis of transport approaches, where
further details of the explicit calculations can be found. Since in the
present study we want to explore the possible consequences of a
dropping $\eta$ mass at finite nuclear density we reparameterize the
$\eta$ photoproduction on a proton as
\begin{equation} \label{gam1}
\sigma_{\gamma p \rightarrow \eta p} (\sqrt{s}) = a (1 - x)^b \
\exp\left(-(\sqrt{s} - \sqrt{s_0})^2/c\right)
\end{equation}
with $x = \sqrt{s_0}/\sqrt{s}, a =$ 161.5 $\mu b, b$ = 0.571 and $c$ =
8.9 10$^{-3}$~GeV$^2$, which provides a good description of the experimental
data from Refs.~\cite{exp1,exp2}. The $\eta$ photoproduction on a
neutron is found experimentally to be 2/3 as that on a proton;
we adopt the same ratio
for our calculations.  The parametrization (\ref{gam1}) as a function
of the invariant energy above threshold $\sqrt{s} - \sqrt{s_0}$ now
allows to simulate $\eta$ photoproduction with a dropping $\eta$ mass
according to Eq.~(\ref{etamass}). We note, that the parametrization
(\ref{gam1}) assumes the photoproduction to be dominated by phase space or
the invariant energy above threshold; this assumption still has to be
controlled by microscopic in-medium calculations on $\eta$ photoproduction
and here only serves as an estimate.

The results of our transport calculations for $\eta$ production on
$^{40}$Ca for photon energies from 600 - 780 MeV are displayed in
Fig.~\ref{Figgam} in comparison to the data from Ref.~\cite{Roebig}.
The full line corresponds to the bare $\eta$ mass whereas the dashed
line is obtained for a dropping mass according to Eq.~(\ref{etamass})
with $\alpha$ = 0.18.
The excitation function clearly indicates that the $\eta$ properties do
not change much in the medium except for a strong $\eta$ absorption
which is about 65 \% in case of $^{40}$Ca (cf. Ref.~\cite{Effe}); the
dropping $\eta$-mass scenario leads to enhanced $\eta$ cross sections
especially at low photon energy which are not observed experimentally.
Thus also $\eta$ photoproduction data exclude a dramatic change of the
$\eta$ properties in the medium as possibly 'requested' by the DLS data
(cf.  Fig.~\ref{FigBVfr}).

\section{Anisotropy of the $\eta$ angular distribution}

As proposed by the DLS collaboration \cite{DLSnew} the inconsistency
between the DLS and TAPS data might be due to differences in the
experimental acceptance -- mid-rapidity for TAPS versus forward
rapidity for DLS -- and a strong anisotropy of the $\eta$ angular
distribution.  The TAPS group calculated the total $\eta$ cross
section from a model where mesons are emitted isotropically from a
thermal source at $y_{cm}$.  Using the mid-rapidity TAPS data and a
simple thermodynamical model the DLS collaboration reproduces the
result from our previous  calculation \cite{BCMas96}.

In Fig.~\ref{Figecos} we, therefore,  present the result of our
transport calculation for the angular distribution of $\eta$ mesons for
Ca~+~Ca at 1.0~A$\cdot$GeV.  Mainly due to rescattering in matter the
$\eta$ angular distribution is anisotropic; however, this anisotropy is
not very pronounced.  As was estimated by the TAPS collaboration, the
implementation of this anisotropy gives at maximum a 20\% correction
for the total $\eta$ cross section \cite{TAPSan}. Thus, the anisotropy
effect according to our transport calculations
cannot explain an enhancement by a factor of 3-7 in the new DLS dilepton data.

\section{Summary}

On the basis of the covariant transport approach HSD~\cite{Ehehalt} we
have studied dilepton production and $m_T$ scaling in C~+~C and Ca~+~Ca
collisions at BEVALAC energies.  We have found that the $m_T$ scaling
of $\pi^0$ and $\eta$ observed by the TAPS collaboration is reproduced
by the transport calculation even for the small system C~+~C when
incorporating no medium effects for the pions and etas.

Various contributions are taken into account for dilepton production:
the Dalitz-decays of $\Delta, N^*$ resonances and $\pi^0, \eta, \omega$
mesons, proton-neutron bremsstrahlung, $\pi N$ bremsstrahlung as well
as the direct dilepton decays of the vector mesons $\rho$ and $\omega$.
It was shown that the new DLS data \cite{DLSnew} are underestimated by
a factor of 6-7 in the transport calculations for invariant mass
$0.15\le M\le 0.4$~GeV when no medium effects are involved as in our
previous calculations~\cite{BCMas96}.

Including the $\rho$ meson modifications in the medium (according to
the spectral functions from Refs.~\cite{RappNPA,Peters}) slightly
improves the agreement with the data, but the discrepancy is still
about a factor of 3.  This also holds for the thermodynamical
approach \cite {RappNPA} where an off-shell propagation of pions is
taken into account.  A simple dropping $\eta$ mass scheme together with
the full $\rho$ spectral function can lead to dilepton spectra that are
in a good agreement with the DLS data. However, the $m_T$-scaling of
$\pi^0$'s and $\eta$'s observed by the TAPS collaboration cannot be
reproduced within this dropping $\eta$ mass scenario anymore. The
dropping mass
scenario also is clearly incompatible with the $\eta$ photoproduction
data on nuclei. Also the
anisotropy of the $\eta$ angular distribution due to rescattering
effects in matter - as obtained from our transport calculation -
is much too small to explain the enhancement of the
dilepton yield at $0.15\le M\le 0.4$~GeV as observed by the DLS
collaboration.

\acknowledgements
The authors are grateful for many helpful discussions with
M.~Appenheimer, R.~Averbeck, G.E.~Brown, R.~Holzmann, C.~M.~Ko,
H.~Lenske, V.~Metag, A.~Sibirtsev, V.D.~Toneev, P.~Vogt and
Th.~Weidmann.  They especially like to thank U. Mosel for valuable
suggestions and a careful reading of the manuscript. Furthermore, they
are indebted to M.~Urban for providing the results on the finite
3-momentum dependence in the two-pion selfenergy and to M. Post and W.
Peters for the $\rho$ spectral function from Ref.~\cite{Peters} prior
to publication.  One of us (RR) acknowledges support from the
Alexander-von-Humboldt foundation as a Feodor-Lynen fellow. This work
was supported in part by the U.S. department of energy under contract
No. DE-FG02-88ER40388 and a grant from the National Science Foundation,
NSF PHY 94-21309.

\newpage

\begin{figure}[h]
\caption{The calculated transverse-mass spectra of $\pi^0$ and $\eta$
mesons in comparison with the TAPS data. The upper part shows the $m_T$
spectra for $\pi^0$'s (dashed histogram) and $\eta$'s (solid histogram)
for C~+~C at 1.0~A$\cdot$GeV in the rapidity interval $0.42 \le y \le
0.74$ and at 2.0~A$\cdot$GeV for $0.8 \le y \le 1.08$.  The
experimental data -- open circles and solid squares correspond to
$\pi^0$ and $\eta$ mesons, respectively -- are taken from
Ref.~\protect\cite{TAPS-CC}. The theoretical results as well as the
experimental data at 2.0 A$\cdot$GeV are multiplied by a factor of
$10^2$.  The middle part corresponds to Ca~+~Ca at 1.0~A$\cdot$GeV  for
$0.48 \le y \le 0.88$ (multiplied by $10^{-1}$) and at 2.0~A$\cdot$GeV
for $0.8 \le y \le 1.1$ in comparison with the data from
Ref.~\protect\cite{TAPS-CaNi}.  The lower part shows the calculated
$m_T$ spectra for Ni~+~Ni at 1.93~A$\cdot$GeV at rapidities $0.8 \le y
\le 1.1$ in comparison with the data from
Ref.~\protect\cite{TAPS-CaNi}. }
\label{Figmt}
\end{figure}

\begin{figure}[t]
\caption{The (negative) imaginary part of the $\rho$ propagator
averaged over the longitudinal and transverse components (where the latter
contain the correction Eq.~(\ref{KLZ})) as a function
of the invariant mass $M$ and the momentum $q$ for baryon densities of
0, 1 $\rho_0$, 2 $\rho_0$, and 3 $\rho_0$ and temperature $T = 70$~MeV
in the extended approach of Ref. \protect\cite{RappNPA}.
Note the different absolute scales in the individual figures.}
\label{FigRapp}
\end{figure}

\begin{figure}[t]
\caption{The (negative) imaginary part of the $\rho$ propagator
averaged over the longitudinal and transverse components as a function
of the invariant mass $M$ and the momentum $q$ for baryon densities of
0, 1 $\rho_0$, 2 $\rho_0$, and 3 $\rho_0$ at temperature $T = 0$~MeV
in the approach of Ref. \protect\cite{Peters}.
Note the different absolute scales in the individual figures. }
\label{FigPeters}
\end{figure}

\begin{figure}[h]
\vspace*{5mm}
\caption{The dilepton spectra (full solid line) for Ca~+~Ca (upper
part) and C~+~C (lower part) at 1.0~A$\cdot$GeV calculated with the
'free' $\rho$ spectral function including the DLS acceptance filter
(version 4.1) as well a mass resolution $\Delta M/M=10\%$ in comparison
with the data from Ref.~\protect\cite{DLSnew}.  The thin lines indicate
the individual contributions from the different production channels
including the DLS acceptance and mass resolution; {\it i.e.}
starting from low $M$: Dalitz decay $\pi^0 \to \gamma e^+ e^-$ (dashed
line), $\eta \to \gamma e^+ e^-$ (dotted line), $\Delta \to N e^+ e^-$
(dashed line), $\omega \to \pi^0 e^+ e^-$ (dot-dashed line), $N^* \to N
e^+ e^-$ (dotted line), proton-neutron bremsstrahlung (dot-dashed
line), $\pi N$ bremsstrahlung (dot-dot-dashed line); for $M \approx $
0.8 GeV:  $\omega \to e^+e^-$ (dot-dashed line), $\rho^0 \to e^+e^-$
(dashed line), $\pi^+ \pi^- \to \rho \to e^+e^-$ (dot-dashed line).}
\label{FigBVfr}
\end{figure}

\begin{figure}[h]
\caption{The dilepton spectra (full solid line) for Ca~+~Ca (upper
part) and C~+~C (lower part) at 1.0~A$\cdot$GeV calculated with the
full $\rho$ spectral function in the extended approach of
Ref. \protect\cite{RappNPA} in comparison with the data from
Ref.~\protect\cite{DLSnew}.  The assignment of the individual
contributions is the same as in Fig.~\protect\ref{FigBVfr}.}
\label{FigBVsf}
\end{figure}

\begin{figure}[h]
\caption{The dilepton spectra (full solid line) for Ca~+~Ca (upper
part) and C~+~C (lower part) at 1.0~A$\cdot$GeV calculated with the
full $\rho$ spectral function in the extended approach of
Ref. \protect\cite{Peters} in comparison with the data from
Ref.~\protect\cite{DLSnew}.  The assignment of the individual
contributions is the same as in Fig.~\protect\ref{FigBVfr}.}
\label{figp}
\end{figure}

\begin{figure}[h]
\caption{The dilepton spectra for Ca~+~Ca at 1.0~A$\cdot$GeV calculated
within the thermodynamical approach \protect\cite{RappNPA} (solid lines)
and within the transport model (dashed lines).}
\label{FigBVtermo}
\end{figure}

\begin{figure}[h]
\caption{The dilepton spectra (full solid line) for Ca~+~Ca (upper
part) and C~+~C (lower part) at 1.0~A$\cdot$GeV calculated with the
full $\rho$ spectral function in the extended approach of Ref.
\protect\cite{RappNPA} for the dropping $\eta$ mass scenario
in comparison with the data from Ref.~\protect\cite{DLSnew}.
The assignment of the individual contributions is the same as in
Fig.~\protect\ref{FigBVfr}.}
\label{FigBVet}
\end{figure}

\begin{figure}
\caption{The transverse-mass spectra of $\pi^0$ and $\eta$ mesons for
Ca~+~Ca in the rapidity interval $0.48 \le y \le 0.88$ (upper part)
and C~+~C for $0.42 \le y
\le 0.74$  (lower part) at 1.0~A$\cdot$GeV. The assignment is the same
as in Fig.~\protect\ref{Figmt}.  The dot-dashed histograms correspond
to the $\eta$ $m_T$-spectra obtained within the dropping $\eta$ mass
scenario.}
\label{Figmted}
\end{figure}

\begin{figure}
\caption{The calculated $\eta$ photoproduction cross section on
$^{40}$Ca as a function of the photon energy $E_\gamma$ in comparison
to the data from \protect\cite{Roebig}. The solid histogram displays
our results for a bare $\eta$ mass whereas the dashed histogram
corresponds to the dropping $\eta$ mass scenario
(\protect\ref{etamass}).}
\label{Figgam}
\end{figure}

\begin{figure}
\caption{The calculated angular distribution of $\eta$ mesons for Ca~+~Ca
at 1.0~A$\cdot$GeV in the center-of-mass system. }
\label{Figecos}
\end{figure}

\newpage
\psfig{figure=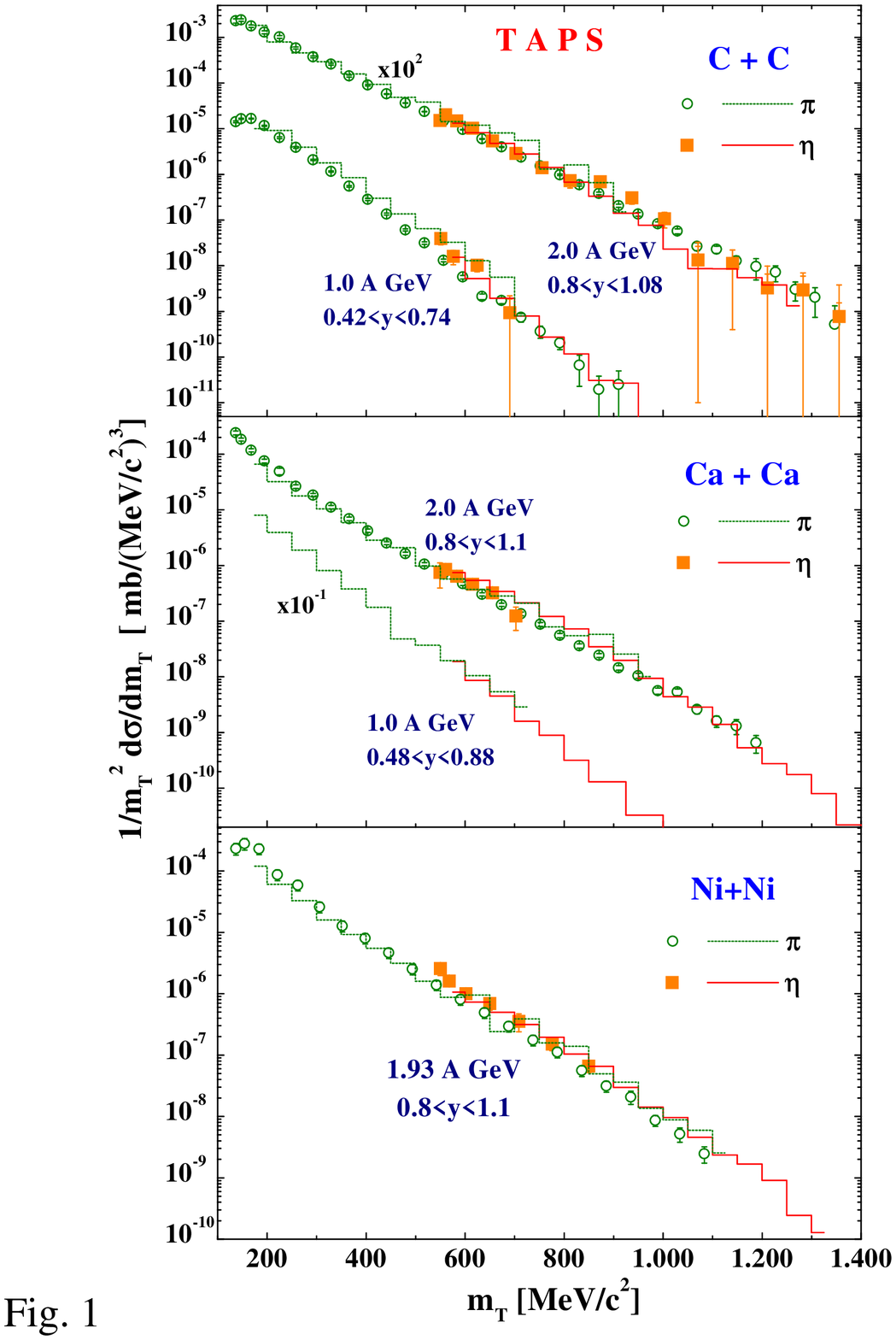,width=15cm,height=22cm}
\psfig{figure=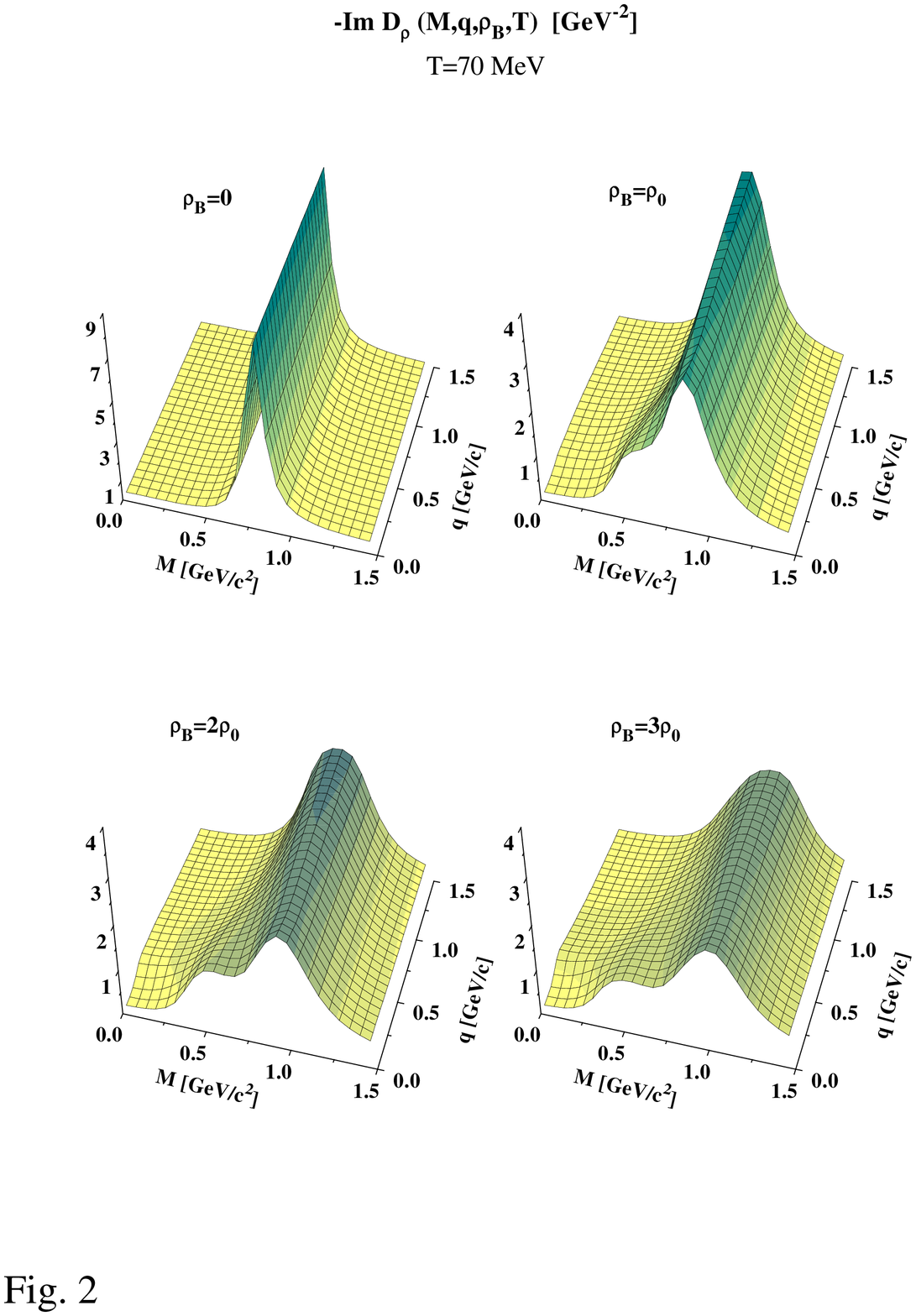,width=15cm,height=22cm}
\psfig{figure=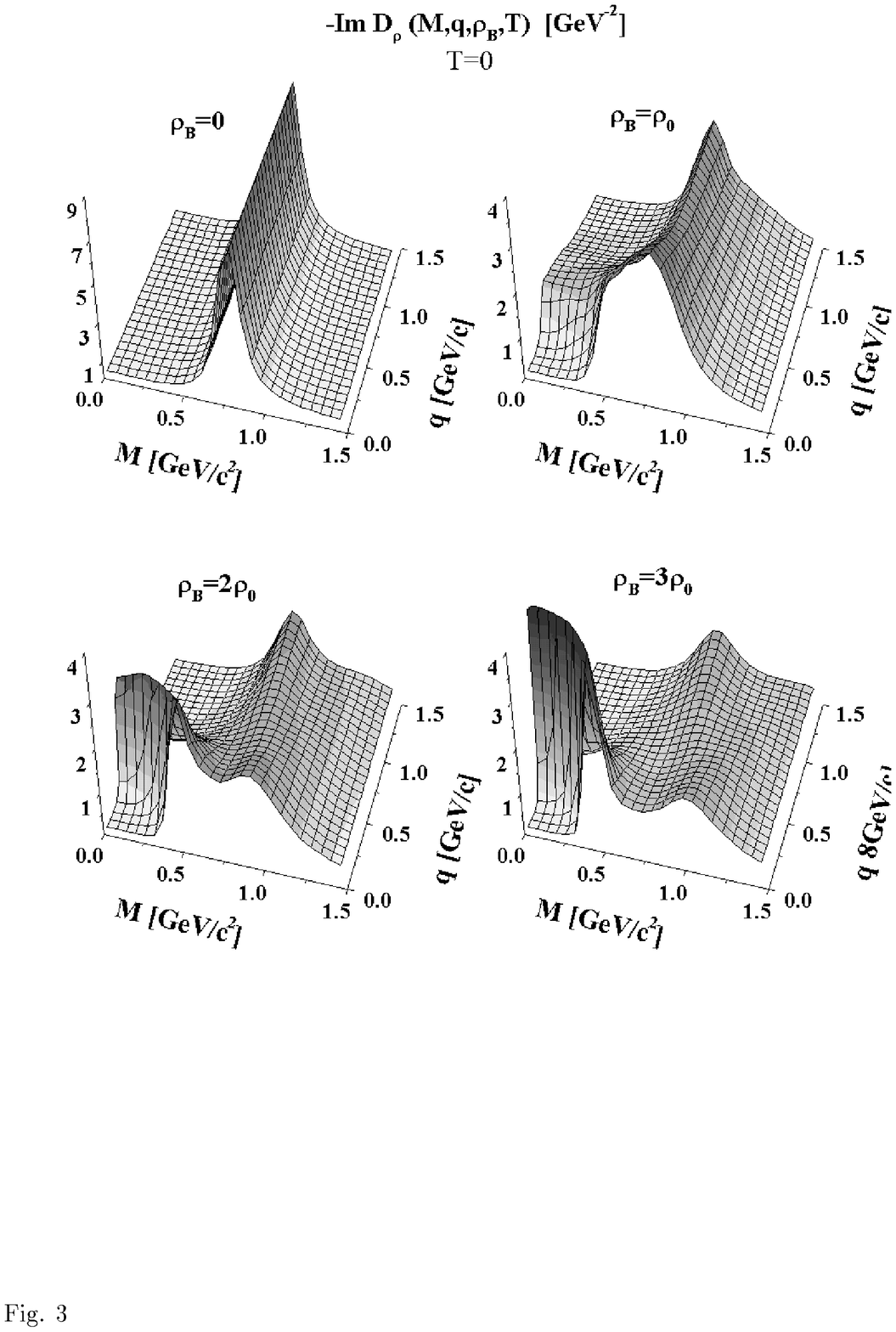,width=15cm,height=22cm}
\psfig{figure=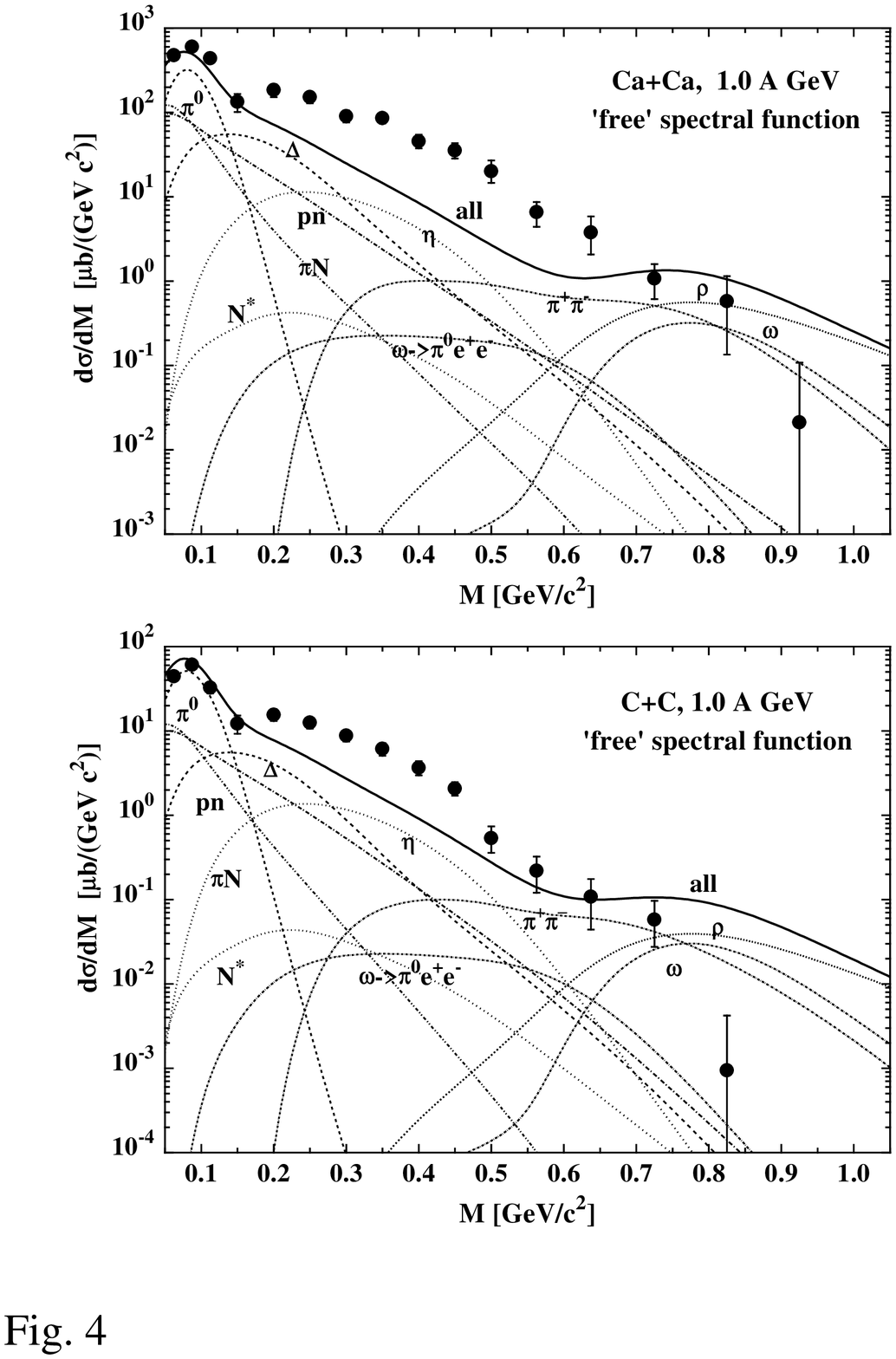,width=15cm,height=22cm}
\psfig{figure=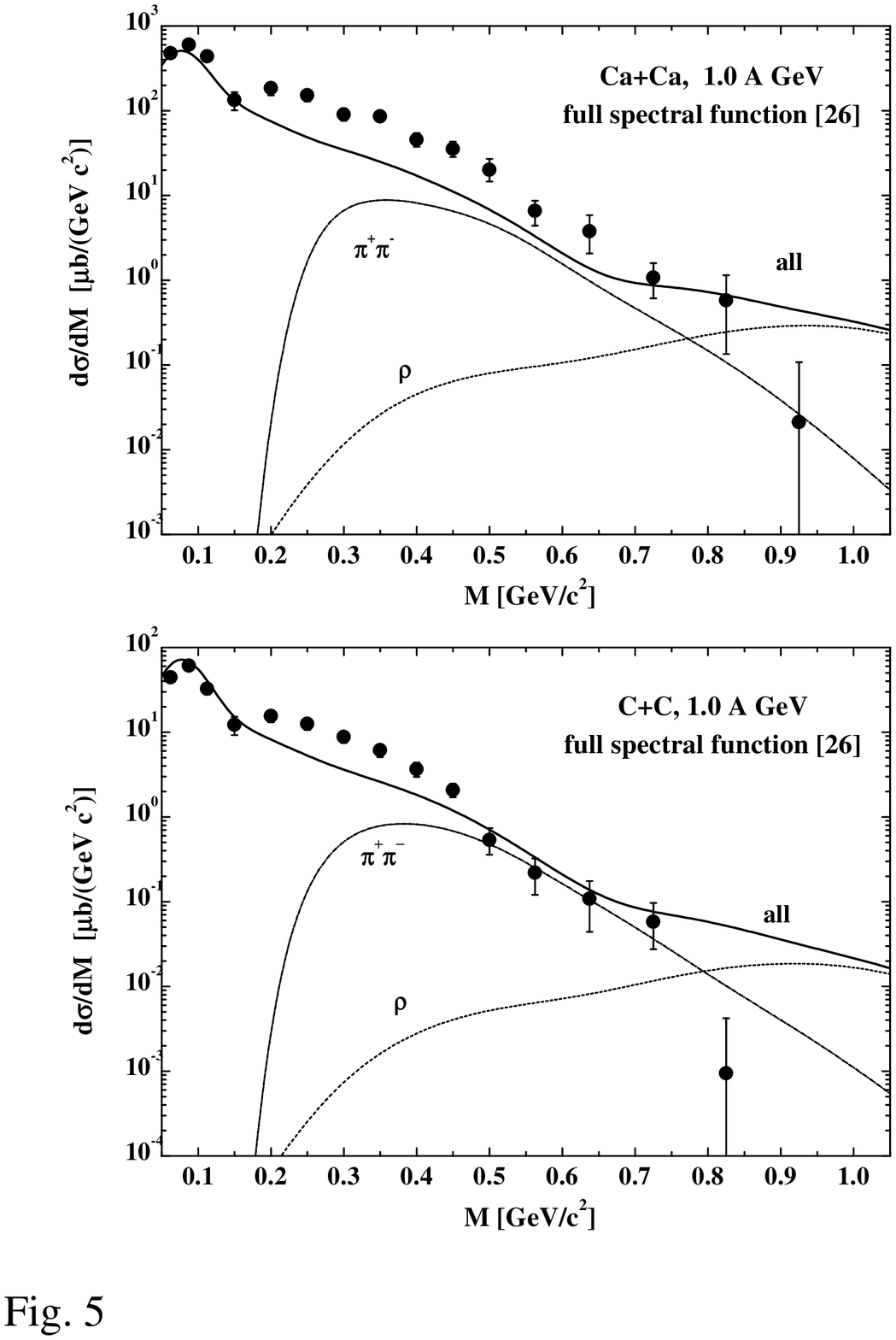,width=15cm,height=22cm}
\psfig{figure=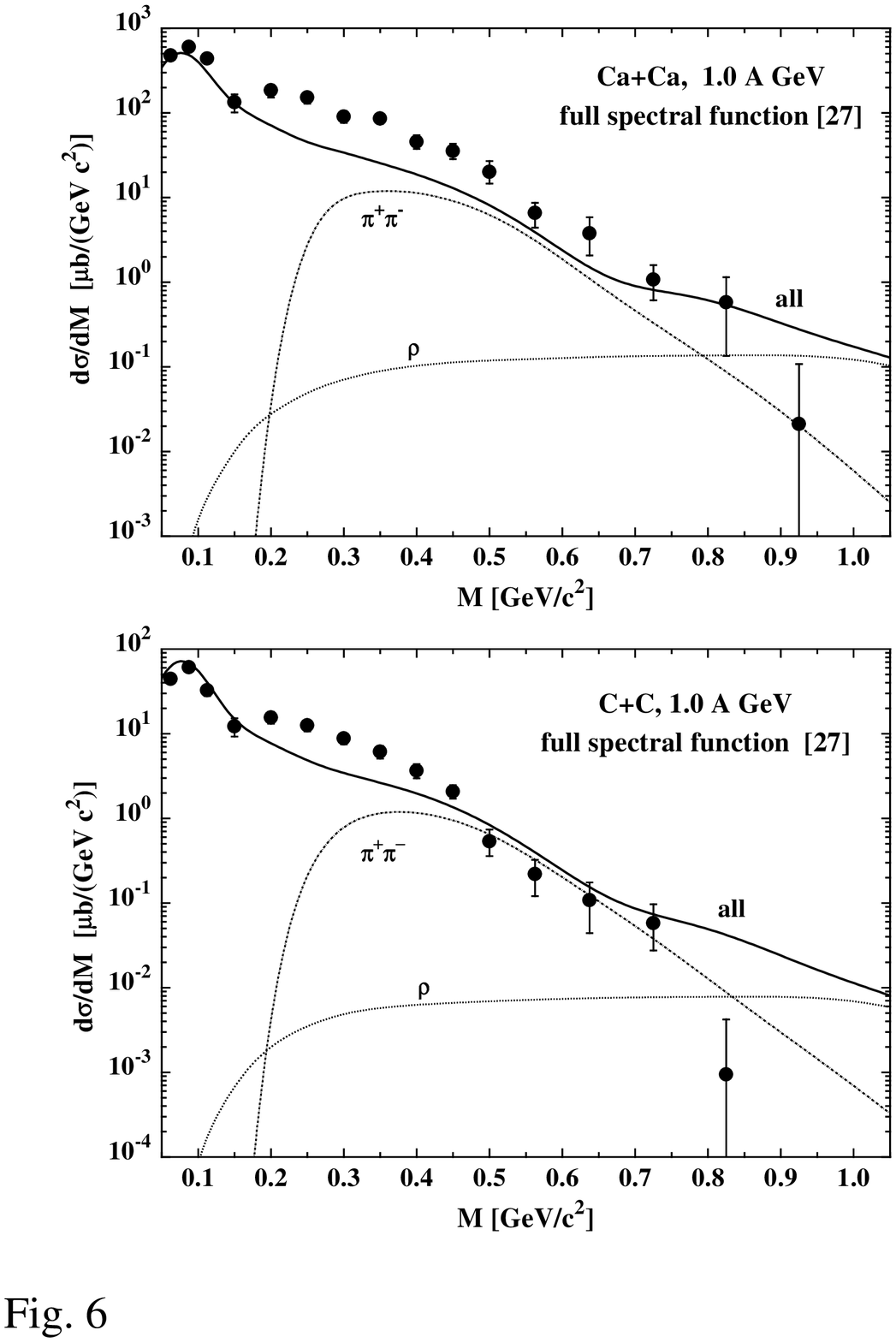,width=15cm,height=22cm}
\psfig{figure=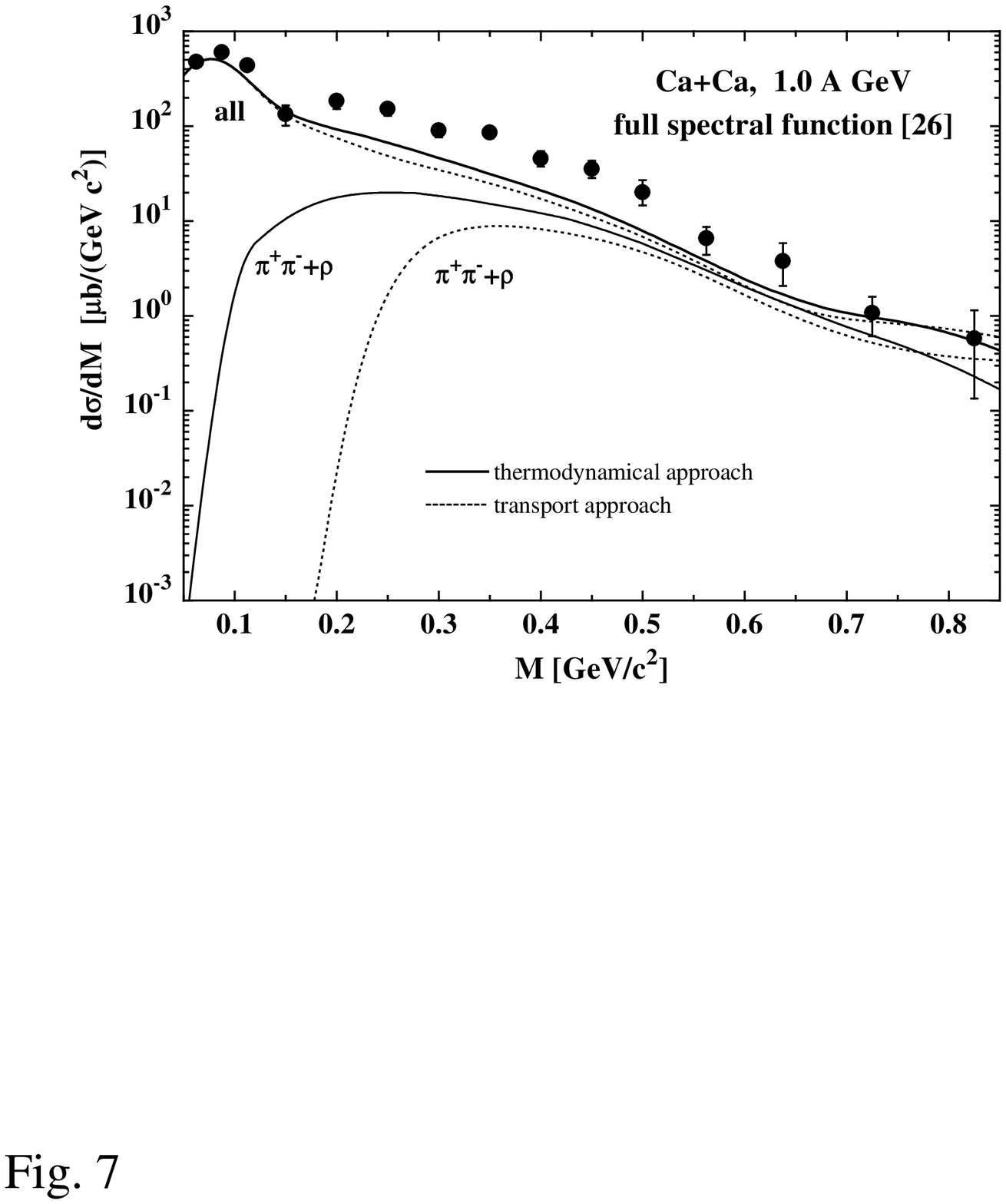,width=15cm,height=22cm}
\psfig{figure=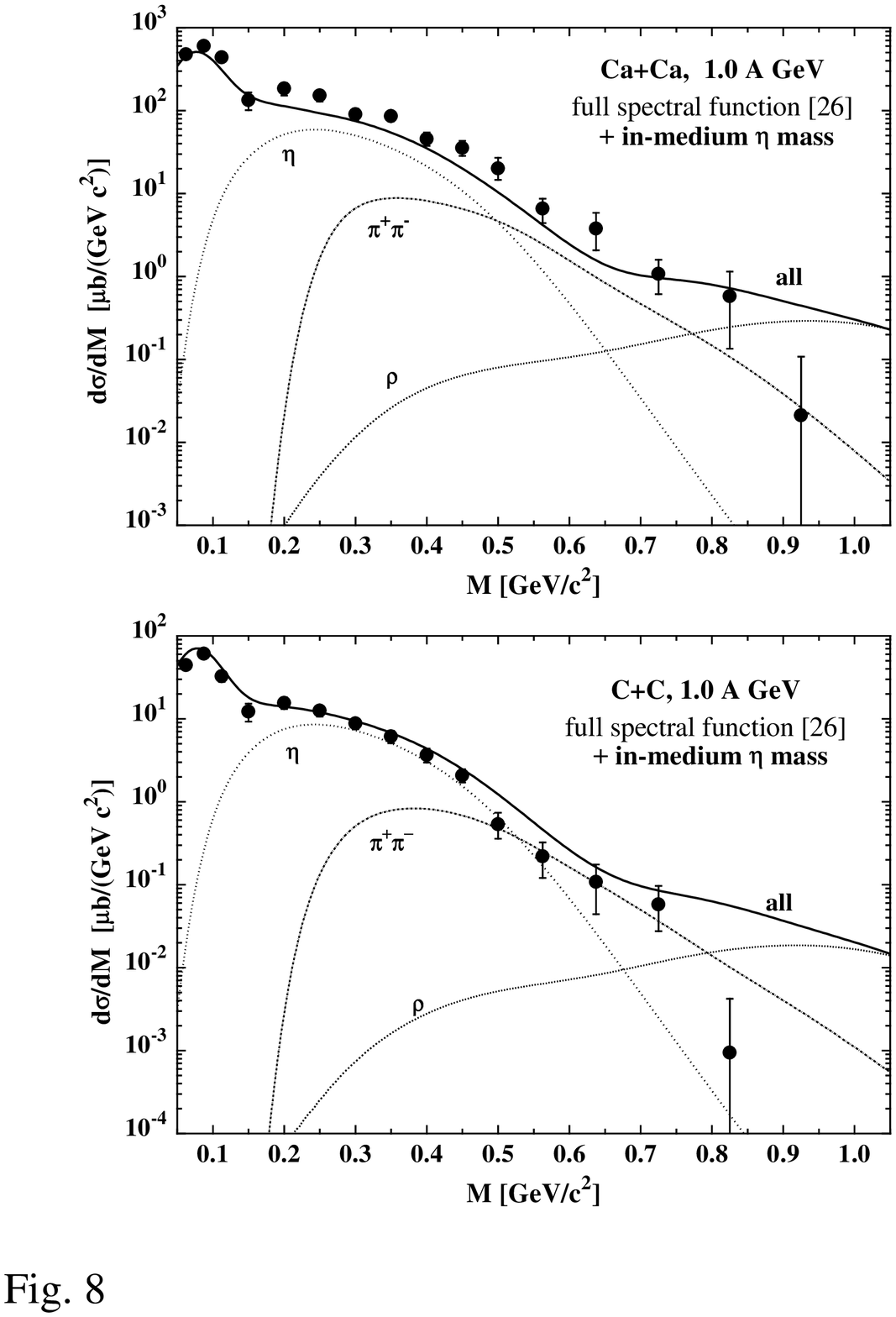,width=15cm,height=22cm}
\psfig{figure=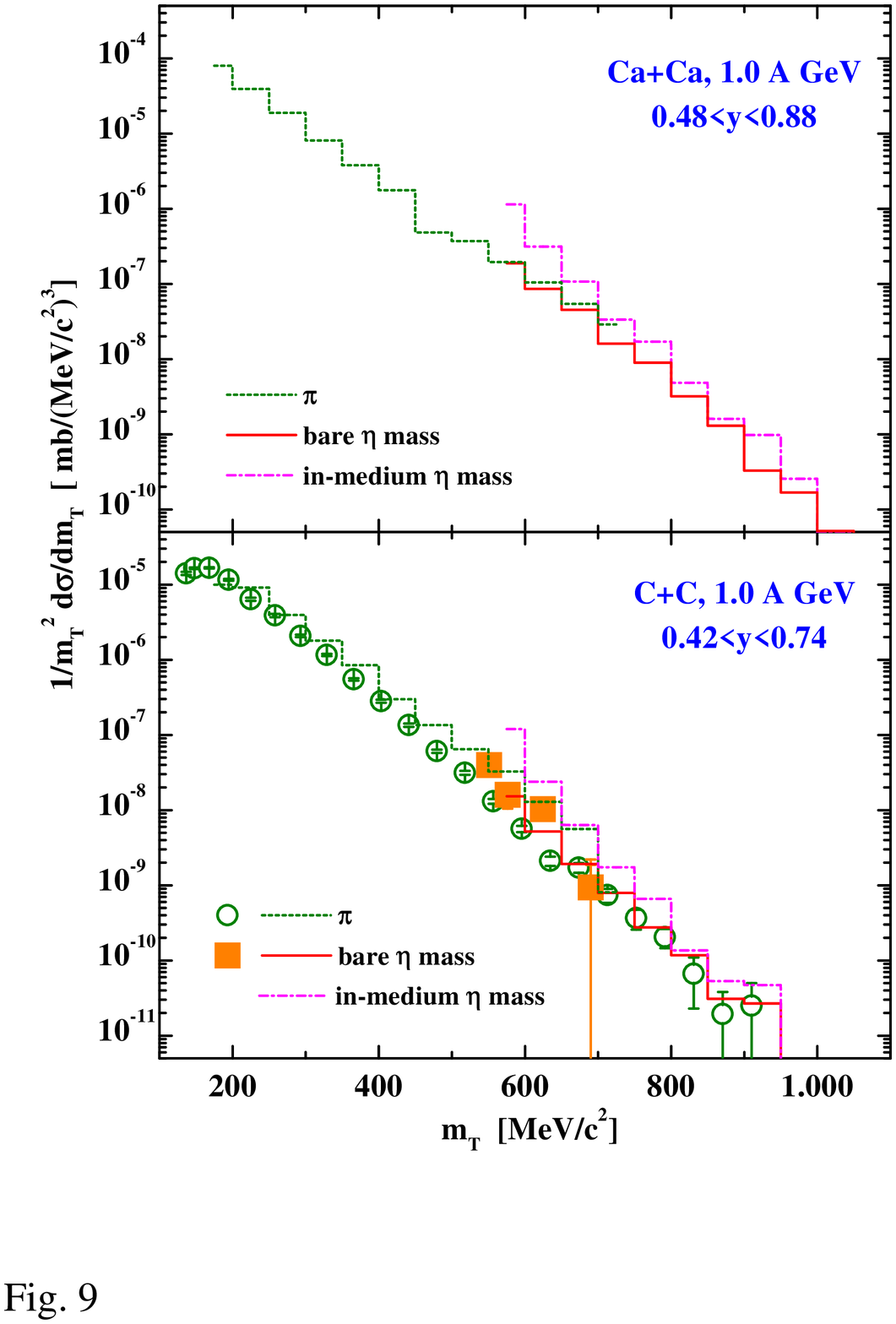,width=15cm,height=22cm}
\psfig{figure=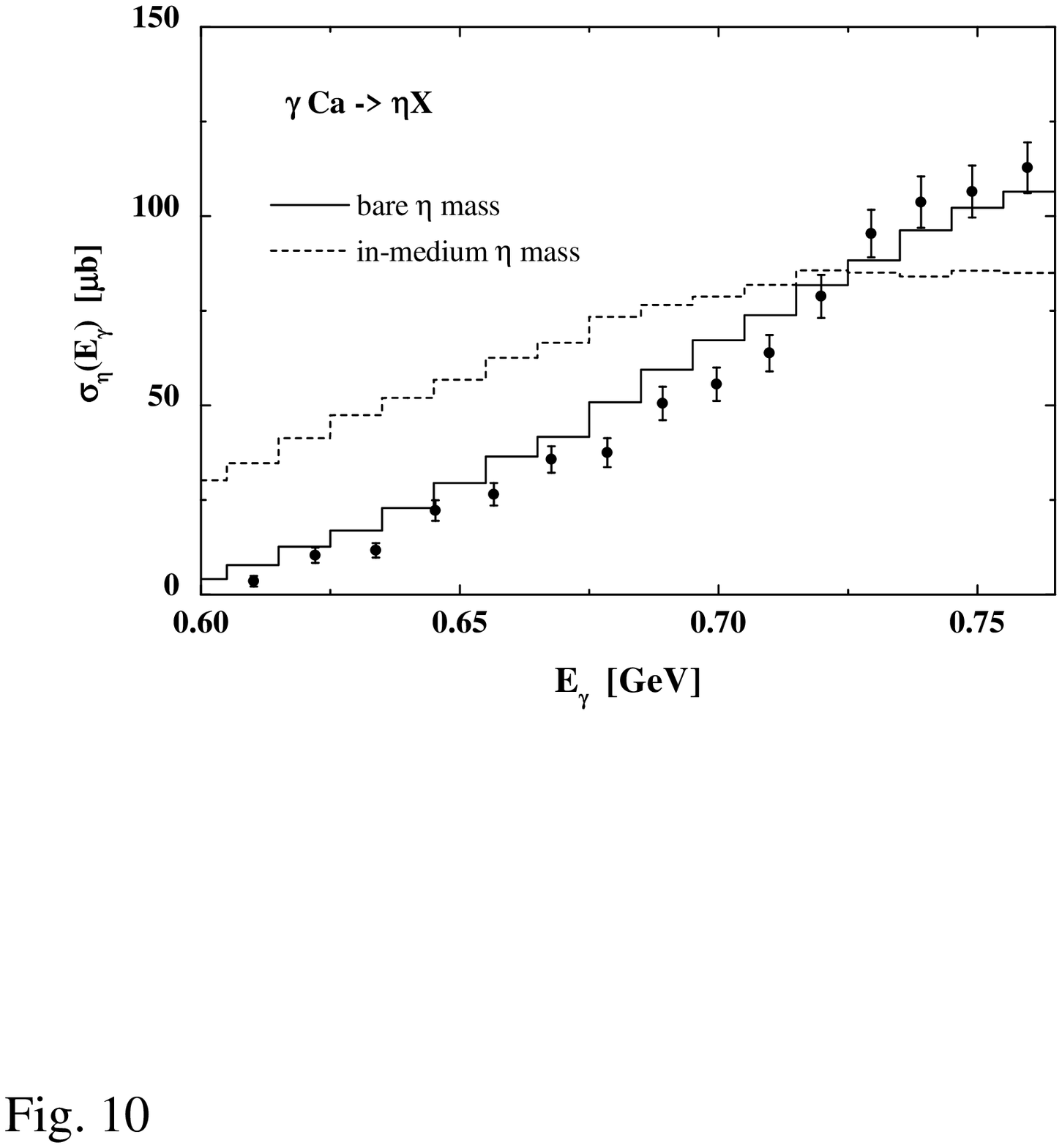,width=15cm,height=22cm}
\psfig{figure=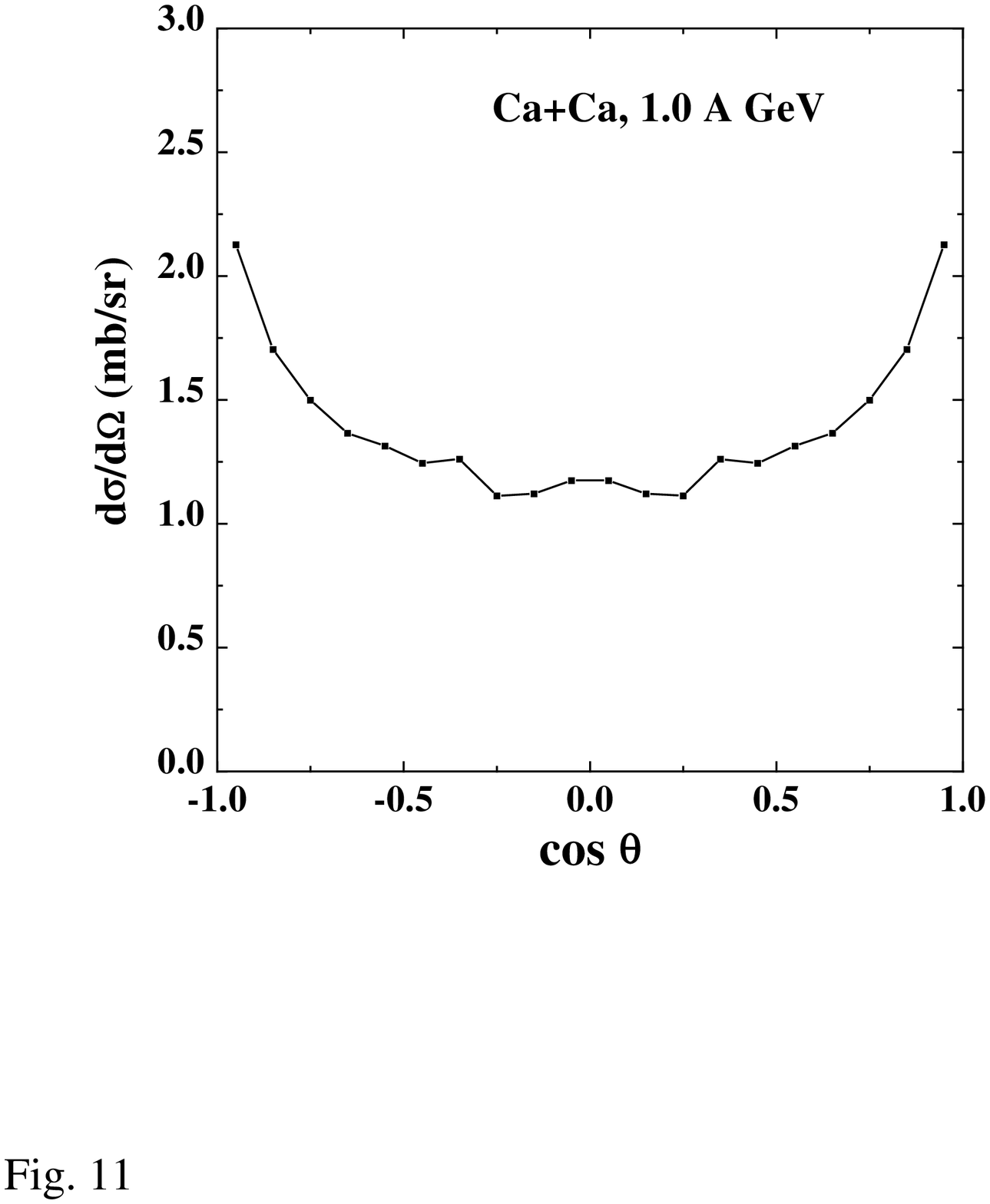,width=15cm,height=22cm}

\end{document}